\documentclass[sigconf,authorversion]{acmart}

\usepackage{balance}

\def\BibTeX{{\rm B\kern-.05em{\sc i\kern-.025em b}\kern-.08emT\kern-.1667em\lower.7ex\hbox{E}\kern-.125emX}}
    
\copyrightyear{2019} 
\acmYear{2019} 
\setcopyright{acmcopyright}
\acmConference[KDD '19]{The 25th ACM SIGKDD Conference on Knowledge Discovery and Data Mining}{August 4--8, 2019}{Anchorage, AK, USA}
\acmBooktitle{The 25th ACM SIGKDD Conference on Knowledge Discovery and Data Mining (KDD '19), August 4--8, 2019, Anchorage, AK, USA}
\acmPrice{15.00}
\acmDOI{10.1145/3292500.3330769}
\acmISBN{978-1-4503-6201-6/19/08}

\usepackage{graphicx}
\usepackage[flushleft]{threeparttable}
\usepackage{adjustbox}

\usepackage{amsmath, amsthm, amssymb, mathtools}
\usepackage{color}
\usepackage{enumitem}

\newcommand{\overbar}[1]{\mkern 1.5mu\overline{\mkern-1.5mu#1\mkern-1.5mu}\mkern 1.5mu}

\newcommand\independent{\protect\mathpalette{\protect\independenT}{\perp}}
\def\independenT#1#2{\mathrel{\rlap{$#1#2$}\mkern2mu{#1#2}}}

\DeclareMathOperator*{\argmin}{arg\,min}

\DeclarePairedDelimiter{\norm}{\lVert}{\rVert}

\theoremstyle{plain}
\newtheorem{theorem}{Theorem}
\newtheorem{lemma}{Lemma}

\theoremstyle{definition}
\newtheorem{definition}{Definition}
\newtheorem{assumption}{Assumption}

\newtheorem{example}{Example} %

\theoremstyle{remark} %

\usepackage{algorithm}
\usepackage{algpseudocode}
\algnewcommand\algorithmicinput{\textbf{Input}}
\algnewcommand\algorithmicoutput{\textbf{Output}}
\algnewcommand\Input{\item[\algorithmicinput]}%
\algnewcommand\Output{\item[\algorithmicoutput]}%

\acmSubmissionID{ads1688o}

\begin{document}

%
\title[Causal Mediation Analysis for A/B Tests]{The Identification and Estimation of Direct and Indirect Effects in A/B Tests through Causal Mediation Analysis}

%

\author{Xuan Yin}
\affiliation{%
  \institution{Etsy, Inc.}
  \streetaddress{117 Adams St}
  \city{Brooklyn}
  \state{New York}
  \postcode{11201}
  }
\email{xyin@etsy.com}

\author{Liangjie Hong}
\affiliation{%
  \institution{Etsy, Inc.}
  \streetaddress{117 Adams St}
  \city{Brooklyn}
  \state{New York}
  \postcode{11201}
  }
\email{lhong@etsy.com}

%
\renewcommand{\shortauthors}{Yin and Hong}

%
\begin{abstract}
E-commerce companies have a number of online products, such as organic search, sponsored search, and recommendation modules, to fulfill customer needs. Although each of these products provides a unique opportunity for users to interact with a portion of the overall inventory, they are all similar channels for users and compete for limited time and monetary budgets of users. To optimize users' overall experiences on an E-commerce platform, instead of understanding and improving different products separately, it is important to gain insights into the evidence that a change in one product would induce users to change their behaviors in others, which may be due to the fact that these products are functionally similar. In this paper, we introduce causal mediation analysis as a formal statistical tool to reveal the underlying causal mechanisms. Existing literature provides little guidance on cases where multiple unmeasured causally-dependent mediators exist, which are common in A/B tests.  We seek a novel approach to identify in those scenarios direct and indirect effects of the treatment. In the end, we demonstrate the effectiveness of the proposed method in data from Etsy's real A/B tests and shed lights on complex relationships between different products.
\end{abstract}

%
%


%
\keywords{A/B test; causal inference; mediation analysis; potential outcome; experiment; online product}

%

%
\maketitle

\section{Introduction}\label{sec:introduction}
Nowadays an internet company typically has a wide range of online products, which are usually functionally similar, to fulfill customer needs. For example, in E-commerce platforms, organic search is a powerful tool for buyers to find interesting listings. Meanwhile, these platforms (e.g., Etsy, eBay) commonly offer opportunities for sellers to promote their listings in search results through advertising, also known as promoted listings. These offerings naturally compete with the organic search to attract users' attention to the same set of purchase intents. In addition, E-commerce platforms have recommendation modules to inspire users based on their past behaviors. Although each of these products provides a unique opportunity for users to interact with a portion of the overall inventory, they are all similar channels for users and contest the limited time and monetary budgets of users.

To optimize users' overall experiences on an E-commerce platform, instead of understanding and improving these products separately, it is important to gain insights into the evidence that a change in one product would induce users to change their behaviors in others, which may be due to the fact that those products are functionally similar.  Below are two motivating examples from Etsy:
\begin{example}\label{example1}
In A/B tests of certain recommendation modules, we sometimes observe statistically significant lifts in the number of clicks on recommendation modules while statistically significant decreases in the number of clicks on organic search results, as well as statistically insignificant lifts in the sitewide conversion rate. 
\end{example}
\begin{example}\label{example2}
In A/B tests of promoted listings, we sometimes observe statistically significant lifts in key metrics of promoted listings (e.g., Click-Through-Rate ({\tt CTR}), the number of clicks, the advertising revenue) while statistically significant decreases in the number of clicks on organic search results, as well as statistically insignificant decreases in the sitewide conversion rate.  
\end{example}
These two examples indicate the possible competitions among products.  To distinguish from the change in user behaviors in the tested product, we refer to those changes in user behaviors in other products as \textit{induced changes}.

The induced changes challenge decision makers and product engineers when deciding the launch of new products through A/B tests.  Although the company typically uses the sitewide conversion rate as an indicator of the marketplace to make decisions, in Example~\ref{example1}, it would be difficult to decide to ramp down the new version of the recommendation module given its improved user engagement.  Moreover, if the change in the recommendation module should have led to a significant lift of the sitewide conversion rate but its potential contribution is offset by the induced reduction of user engagement in organic search, it is desirable to launch the new version and to improve the organic search later.  In Example~\ref{example2}, if the company takes improving the organic search as a priority over improving other products, it would be difficult to decide to launch the new version of promoted listings given the impaired organic search performance.  Furthermore, decision makers would like to know how much such degradation could be tolerated. 

These challenges ask for separating the effects of the induced change from the direct effects of the change in the tested product (e.g., how the induced decrease in user engagement of organic search impacts the sitewide conversion rate.) Unfortunately, such an evaluation cannot be directly derived from A/B tests because these tests can only assess effects of the change in a tested product (e.g., how the change in recommendation modules or promoted listings impacts the sitewide conversion).

In this paper, we utilize a formal statistical framework \textit{Causal Mediation Analysis} ({\tt CMA}) to estimate \textit{causal} effects of the induced changes.  {\tt CMA} is widely adopted to reveal the underlying causal mechanism in randomized experiments of various disciplines.  Here, the causal mechanism is referred to as a process that a treatment affects the outcome through some intermediate variables that can be referred to as \textit{mediators}.  In Example~\ref{example1}, the change in a recommendation module increases clicks on the recommendation module but decreases clicks on organic search results, and the lift of overall conversion rate is insignificant.  A possible mechanism could be, an improved recommendation module successfully suggests listings that satisfy users' needs so that users do not need organic search as much as usual, and the direct positive effect of the improvement of the recommendation module upon the sitewide conversion is offset by the negative effect of the induced reduction in user engagement of search so that the total effect is insignificant.  {\tt CMA} can split the total effect of the change in the recommendation module upon conversion (i.e., treatment effect) into a direct component (i.e., direct effect) and an indirect component (i.e., causal mediation effect) where the direct component goes directly from the change to the conversion and the indirect component is transmitted by the organic search to the conversion.

One challenge to conduct {\tt CMA} for A/B tests is that a single intervention can change many causal variables at once, which is often called ``fat hand.''~\citep{Peysakhovich2018LearningVariables}  Those causal variables might be causally-dependent.  Unmeasured causally-dependent mediators could break the sequential ignorability ({\tt SI}) assumption of {\tt CMA} so that it invalidates the identification of direct and indirect effects.  Etsy has hundreds of webpages and modules. In Example~\ref{example1}, the change in the recommendation module can induce users to change their behaviors on many other webpages and modules.  Therefore, if the number of organic search clicks is the mediator of interest, numerous mediators might confound its relationship with the sitewide conversion.  However, it could be too costly to measure user behaviors on every single webpage or module.  We rigorously prove that under certain assumptions, even when multiple unmeasured causally-dependent mediators exist, we can still identify and estimate a generalized direct effect and indirect effect via two linear regression equations.

To summarize, our contributions in this paper include:
\setlist{nolistsep}
\begin{enumerate}
    \item This is the first study to our knowledge that utilizes {\tt CMA} to evaluate the induced change in user behaviors in online services, which cannot be assessed directly from A/B tests. In addition, the average direct effect derived from {\tt CMA} represents the average direct impact of the change in a tested product upon the outcome (e.g., conversion), which could be a better metric to assess the success of the tested product. 
    \item Unmeasured causally-dependent mediators can break the identification assumption ({\tt SI}) in {\tt CMA}. For our purposes, we generalize the definition of direct and indirect effects by explicitly incorporating unmeasured causally-dependent mediators.  Our generalized effects collapse to the direct and indirect effects when there is no unmeasured causally-dependent mediator.  We rigorously prove that under certain assumptions, the generalized effects can be identified, even when multiple unmeasured causally-dependent mediators exist. We further show that the generalized effects can be estimated via two linear regression equations, which are easy to estimate in practice and do not need extra randomization.
    \item We demonstrate the effectiveness of the proposed approach on data from Etsy's real A/B tests and show interesting insights from the analysis.
\end{enumerate}

\section{Literature Review}\label{sec:related_work}
Mediation analysis can be traced back to the seminal paper of \citet{Baron1986TheConsiderations.} in psychology, which supplies a simple multivariate regression framework for applied analysis.  Although the method has been widely employed in various disciplines since then, it has been unclear that the \textit{assumptions} to justify a \textit{causal interpretation} of the estimator until researchers reframe the problem using formal languages of causal inference~\citep{Pearl2001DirectEffects,Robins2003SemanticsEffects,Imai2010IdentificationEffects}.  The core research question is that, what kind of assumption is needed to identify what kind of causal effect.  On one hand, computer scientists and epidemiologists use causal graphical model for mediation analysis and the well-known models include the {\tt NPSEM-IE} of \citet{Pearl2001DirectEffects} and the {\tt FFRCISTG} of \citet{Robins2003SemanticsEffects}.  \citet{Shpitser2013CounterfactualConfounding} identifies the direct effect in a longitudinal setting when unobserved confounder exists.  On the other hand, statisticians extend the potential outcome framework to define and analyze causal mediation effects.  For example, \citet{Imai2010IdentificationEffects} achieves the nonparametric identification of mediation effects of a single mediator under {\tt SI} within potential outcome framework. And \citet{Imai2013IdentificationExperiments} discusses the identification problem of multiple causally-dependent mediators when there is no unmeasured post-treatment covariate.

In this paper, we follow the literature of potential outcome framework to conduct {\tt CMA}.  {\tt CMA} of potential outcome framework is a proven approach to understand the underlying causal mechanism of randomized experiments in applied research areas, such as psychology~\citep{Rucker2011MediationRecommendations}, political science~\citep{Imai2013IdentificationExperiments}, economics~\citep{Heckman2015EconometricInputs}, public policy~\citep{Keele2015IdentifyingAnalysis}, and nursing~\citep{Hertzog2017TrendsPractice}.  Our discussion will be limited to {\tt CMA} of potential outcome framework.

{\tt SI} is the most widely deployed assumption for the identification of causal effects in {\tt CMA}.  The literature explores various identification assumptions for two causally-dependent mediators~\citep{Imai2013IdentificationExperiments,TchetgenTchetgen2014IdentificationExposure}.  However, most studies do not address the identification problem with multiple unmeasured causally-dependent mediators that could break {\tt SI}.  Many of them assume no unmeasured variables when discussing identification and propose various sensitivity analysis to simulate the impacts of unmeasured confounders~\citep{Imai2010IdentificationEffects,Imai2013IdentificationExperiments,Small2012MediationVariables}.

A few studies examine the identification problem when there are unmeasured confounders of the mediator-outcome relationship~\citep{TenHave2007CausalModels,Small2012MediationVariables,Ogburn2012CommentarySmall}.  They use observed baseline covariates, together with treatment assignment, to construct the instrumental variable for the identification of the mediation effects.  The method can suffer from weak identification problem.  Yet, the instrumental variable approach may not be useful for Etsy's A/B tests.  In Etsy, because of our ``guest checkout'' function (i.e., users can make purchases without registering at Etsy) and increasingly stringent privacy protection, it is difficult to obtain users' baseline covariates, such as age, gender, and education level, so that we are unable to construct instrumental variables.  Therefore, we seek a novel approach to better align with Etsy's A/B tests. 

\section{Preliminary}\label{sec:preliminary}
We follow the literature of {\tt CMA} of the potential outcome framework, which is a proven method of causal inference in applied work. Here we use Example~\ref{example1} to clarify the framework.  The exposition below mainly draws on \citet{Rubin2003BasicStudies}, \citet{Imai2010IdentificationEffects}, and \citet{Keele2015IdentifyingAnalysis}.

In an A/B test, the experimental unit is a user, the treatment is the change in the tested product, and the outcome could be, for example, conversion, which is a common business interest.  Let $T_i$ denote the treatment that the user $i$ is assigned.  In Example~\ref{example1}, $T_i=1$ if she was presented with the new version of the recommendation module, and $T_i=0$ if she was presented with the old version.  Here, we focus on binary treatment.  Let $M_i(t)$ denote the potential mediator of the user $i$ under treatment $t$.  In Example~\ref{example1}, $M_i(1)$ is the number of times the user $i$ clicked on organic search results if she was presented with the new version of the recommendation module, and $M_i(0)$ is the number of times she clicked on organic search results if she was presented with the old version.  Only one of the potential mediators can be observed for each user in an A/B test.  The observed mediator can be written as $M_i:=M_i(T_i)=(1-T_i)M_i(0)+T_iM_i(1)$.  Let $Y_i(t, m)$ denote the potential outcome of the treatment $t$ and the mediator $m$.  For example, $Y_i(1, 20)$ means the conversion status of the user $i$ if she was presented with the new version of the recommendation module and clicked on organic search results 20 times.  Again, only one of the potential outcomes can be observed for each user in the A/B test.  The observed outcome can be written as $Y_i:=Y_i(T_i, M_i(T_i))=(1-T_i)Y_i(0, M_i(0))+T_iY_i(1, M_i(1))$.

Define individual treatment effect for the user $i$ as
\begin{align*}
Y_i(1, M_i(1)) - Y_i(0, M_i(0)).
\end{align*}

In Example~\ref{example1}, $Y_i(1, M_i(1))$ represents the conversion status of the user $i$ if she clicked on the organic search results $M_i(1)$ times after being presented with the new version of the recommendation module, and $Y_i(0, M_i(0))$ represents her conversion status if she clicked on the organic search results $M_i(0)$ times after being presented with the old version of the recommendation module.

Define causal mediation effect~\citep{Imai2010IdentificationEffects} for the user $i$ as
\begin{align*}
\text{{\tt  CME}}_i(t) = Y_i(t, M_i(1)) - Y_i(t, M_i(0))
\end{align*}
for $t = 0, 1$, which is also known as the natural indirect effect~\citep{Pearl2001DirectEffects}.  In the literature, $\text{{\tt  CME}}_i(0)$ and $\text{{\tt  CME}}_i(1)$ are termed as the pure indirect effect and total indirect effect, respectively~\citep{Robins2003SemanticsEffects}.

Take $\text{{\tt  CME}}_i(1) = Y_i(1, M_i(1)) - Y_i(1, M_i(0))$ as an example.  The potential outcome $Y_i(1, M_i(0))$ represents the conversion status of the user $i$ if she was presented with the new version of the recommendation module but had the same number of organic search clicks as if she had been presented with the old version of the recommendation module (i.e., $M_i(0)$).  Because the user was presented with the new version of the recommendation module all the time (i.e., $T_i$ is fixed at $1$ in the two potential outcomes), the difference between the two potential outcomes (i.e., $\text{{\tt  CME}}_i(1)$) can only be attributed to the difference between the numbers of organic search clicks, which are induced by the different versions of the recommendation module.  Therefore, $\text{{\tt  CME}}_i(1)$ essentially represents the effect of the induced change in the user's organic search clicks upon her conversion status given she was presented with the new version of the recommendation module all the time, which can be utilized to evaluate the induced change in user behaviors in organic search results regarding conversion.

Similarly, define the direct effect of the treatment~\citep{Imai2010IdentificationEffects} for the user $i$ as
\begin{align*}
\text{{\tt  DE}}_i(t) = Y_i(1, M_i(t)) - Y_i(0, M_i(t))
\end{align*}
for $t = 0, 1$, which is also known as the natural direct effect~\citep{Pearl2001DirectEffects}.

Take $\text{{\tt  DE}}_i(0) = Y_i(1, M_i(0)) - Y_i(0, M_i(0))$ as an example.  Because the user clicked the organic search results $M_i(0)$ times no matter which version of the recommendation module she was presented with, the difference between $Y_i(1, M_i(0))$ and $Y_i(0, M_i(0))$ (i.e., $\text{{\tt  DE}}_i(0)$) can only be attributed to the different versions of the recommendation module.  Therefore, $\text{{\tt  DE}}_i(0)$ essentially represents the treatment effect of the change in the recommendation module that goes directly to the conversion status, which can be utilized to evaluate the change in the recommendation module regarding conversion.

Note that, the individual treatment effect can be decomposed into {\tt  CME} and {\tt  DE}
\begin{align*}
Y_i(1, M_i(1)) - Y_i(0, M_i(0)) = \text{{\tt  CME}}_i(1) + \text{{\tt  DE}}_i(0)=\text{{\tt  CME}}_i(0) + \text{{\tt  DE}}_i(1).
\end{align*}

Causal effects defined at the individual level cannot be identified because only one of the potential outcomes could be observed for each user in an A/B test.  The ``identification'' in this paper means, if we know the data from the entire population, whether we can consistently estimate the parameters of interest, such as causal effects, regression coefficients~\citep{Manski2009IdentificationDecision,Lewbal2018TheEconometrics}.  However, it is possible to identify the \textit{average treatment effect} ({\tt  ATE}, also known as the population average treatment effect), the \textit{average causal mediation effect} ({\tt  ACME}), and the \textit{average direct effect} ({\tt  ADE})
\begin{align*}
\text{{\tt  ATE}} &= \mathbb{E}(Y_i(1, M_i(1))) - \mathbb{E}(Y_i(0, M_i(0))) \\
\text{{\tt  ACME}}(t) &= \mathbb{E}(Y_i(t, M_i(1))) - \mathbb{E}(Y_i(t, M_i(0))) \\
\text{{\tt  ADE}}(t) &= \mathbb{E}(Y_i(1, M_i(t))) - \mathbb{E}(Y_i(0, M_i(t))).
\end{align*}
{\tt  ATE} can be nonparametrically identified under two assumptions in Rubin Causal Model~\citep{Holland1986StatisticsInference}: strong ignorability~\citep{Rosenbaum1983TheEffects} and stable unit-treatment-value assumption (SUTVA).  Strong ignorability asks for $\left\lbrace Y_i(0), Y_i(1)\right\rbrace \independent T_i$ and $0<\mathbb{P}(T_i=t)<1$ for $t=0,1$, which can be guaranteed by random assignment in A/B tests.  SUTVA asks for no multiple versions of treatment and no interference between users, which cannot be guaranteed but are implicitly assumed by our notations.  {\tt  ACME} and {\tt  ADE} can be nonparametrically identified under sequential ignorability ({\tt SI}).
\begin{assumption}[Sequential Ignorability, \citet{Imai2010IdentificationEffects}]
\begin{align}
\left\lbrace Y_i(t',m), M_i(t)\right\rbrace \independent T_i&\text{ and }0<\mathbb{P}(T_i=t)<1 \label{imai_si1}\\
Y_i(t',m) \independent M_i(t) \vert  T_i=t&\text{ and }0<\mathbb{P}(M_i(t)=m \vert  T_i=t)<1\label{imai_si2}
\end{align}
for $t, t'=0,1$, all $m\in\mathcal{M}$.
\end{assumption}
Note that, {\tt SI} incorporates strong ignorability (Condition~\eqref{imai_si1}) as its subset.  Condition~\eqref{imai_si2} is unverifiable.  We will discuss {\tt SI} more in Section~\ref{sec:si_multiplemediators}.  

We also assume that the treatment assignment does not depend on any pre-treatment variables, such as the age, education level, and gender of the user, which is common in A/B tests.  However, even if the treatment assignment depends on some observed pre-treatment variables, all the conclusions hold as long as we add those variables as controls in the analysis.
\section{Model with Multiple Unmeasured Causally-Dependent Mediators}\label{sec:model}
As noted in the literature,  unmeasured causally-dependent mediators can break the {\tt SI} and invalidate the identification of {\tt  ADE} and {\tt  ACME}~\citep{Imai2010IdentificationEffects,Imai2013IdentificationExperiments}.  Etsy has hundreds of webpages and modules.  It could be too costly to measure user behaviors on every webpage or module in practice.  Thus, there could be numerous unmeasured causally-dependent mediators in Etsy's A/B tests.

To solve the problem, we generalize the {\tt SI} and definitions of {\tt  ADE} and {\tt  ACME} by explicitly incorporating multiple unmeasured causally-dependent mediators.  We show that the generalized {\tt  ADE} and {\tt  ACME} ({\tt  GADE} and {\tt  GACME}) are still useful for distinguishing the effect of the induced change in user behaviors from the direct effect of the treatment.  We prove that, assuming the generalized {\tt SI} and linear structural equation model ({\tt  LSEM}) for all measured and unmeasured variables, {\tt  GADE} and {\tt  GACME} can be identified through two linear regression equations even when multiple causally-dependent mediators are unmeasured.  The identification implies {\tt  GADE} and {\tt  GACME} can be easily calculated using the coefficients of the two linear regression equations.  There are well-developed techniques and statistical software packages for the estimation of multiple linear regression equations, which facilitates the implementation of our method in practice.

Note that, it could be difficult in practice to know whether there is any unmeasured causally-dependent mediator among hundreds of webpages and modules.  If yes, {\tt  ADE} and {\tt  ACME} cannot be identified while {\tt  GADE} and {\tt  GACME} can; if no, {\tt  GADE} and {\tt  GACME} collapse to {\tt  ADE} and {\tt  ACME}.  Hence, it is advisable for us to focus on {\tt  GADE} and {\tt  GACME} in practice.

\subsection{Causal Effects}
We firstly use directed acyclic graph (DAG) to introduce causally-dependent upstream and downstream mediators.
\begin{figure}[h]
\centerline{\includegraphics[scale = 0.18]{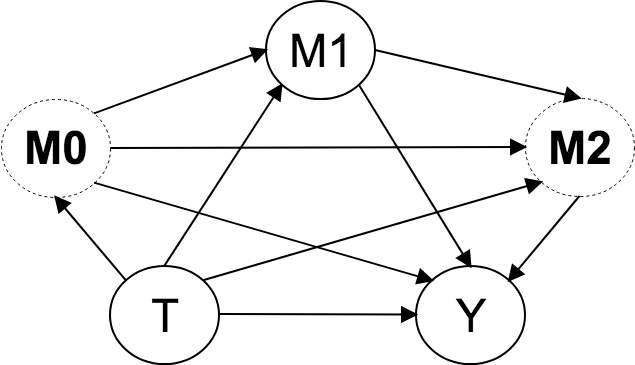}}
\caption{Directed Acyclic Graph (DAG) of Multiple Causally-Dependent Mediators}
\label{fig: mediation2}
\end{figure}

The DAG in Figure~\ref{fig: mediation2} describes all possible causal relationships~\footnote{Following the literature, we do not consider cyclic causality.} among all variables, measured and unmeasured, in an A/B test.  The treatment $T$ impacts the outcome $Y$ partially through mediators $\boldsymbol{M}_0$, $M_1$ and $\boldsymbol{M}_2$.  The group of mediators $\boldsymbol{M}_0$ impacts the group of mediators $\boldsymbol{M}_2$ partially through $M_1$.  The groups of mediators $\boldsymbol{M}_0$ and $\boldsymbol{M}_2$ are upstream and downstream mediators of $M_1$, respectively.  Note that, all the conclusions hold even if there are no upstream mediators or downstream mediators.

With the concepts and notations of upstream and downstream mediators, we can define {\tt  GADE} and {\tt  GACME}.
\begin{definition}[Causal Effects with Multiple Unmeasured Causally-Dependent Mediators]
Define generalized average direct effect ({\tt  GADE}) and generalized average causal mediation effect ({\tt  GACME}) of $M_{i1}$ as
\begin{align*}
\text{{\tt  GADE}}(t)\ =\mathbb{E}[Y_i(1,&\boldsymbol{M}_{i0}(1), M_{i1}(t, \boldsymbol{M}_{i0}(t)),\\
&\boldsymbol{M}_{i2}(1, \boldsymbol{M}_{i0}(1), M_{i1}(t, \boldsymbol{M}_{i0}(t))))]\\
-\mathbb{E}[Y_i(&0, \boldsymbol{M}_{i0}(0), M_{i1}(t, \boldsymbol{M}_{i0}(t)),\\
&\boldsymbol{M}_{i2}(0, \boldsymbol{M}_{i0}(0), M_{i1}(t, \boldsymbol{M}_{i0}(t))))]
\\
\text{{\tt  GACME}}(t)\ = \mathbb{E}[Y_i(t,& \boldsymbol{M}_{i0}(t), M_{i1}(1, \boldsymbol{M}_{i0}(1)),\\
&\boldsymbol{M}_{i2}(t, \boldsymbol{M}_{i0}(t), M_{i1}(1, \boldsymbol{M}_{i0}(1))))]\\
-\mathbb{E}[Y_i(&t, \boldsymbol{M}_{i0}(t), M_{i1}(0, \boldsymbol{M}_{i0}(0)),\\
&\boldsymbol{M}_{i2}(t, \boldsymbol{M}_{i0}(t), M_{i1}(0, \boldsymbol{M}_{i0}(0))))]
\end{align*}
for $t=0,1$.
\end{definition}
It is clear that, without upstream and downstream mediators ($\boldsymbol{M}_{i0}$ and $\boldsymbol{M}_{i2}$), {\tt  GADE} and {\tt  GACME} collapse to {\tt  ADE} and {\tt  ACME}, respectively.  Note that, {\tt  ATE} can be decomposed into {\tt  GADE} and {\tt  GACME} (Appendix~\ref{appendix: eq_proof})
\begin{align}
\text{{\tt  ATE}}
= \text{{\tt  GADE}}&(0)+\text{{\tt  GACME}}(1) = \text{{\tt  GADE}}(1)+\text{{\tt  GACME}}(0).\label{eq: DecomposeATE}
\end{align}
The explanations of {\tt  GADE} and {\tt  GACME} are slightly different from those of {\tt  ADE} and {\tt  ACME}.  In Example~\ref{example1}, $M_{i1}$ represents the number of organic search clicks of the user $i$.  In $\text{{\tt  GADE}}(0)$, to exclude the effect of the induced change in her organic search clicks from {\tt  ATE}, we fix the number of organic search clicks at what it would be if she had been presented with the old version of the recommendation module.  We can see the upstream mediators $\boldsymbol{M}_{i0}$ that are inside the expression of $M_{i1}$ are fixed at $\boldsymbol{M}_{i0}(0)$ so that $M_{i1}$ can be fixed at its potential mediator under the control.  Note that, the upstream mediators $\boldsymbol{M}_{i0}$ that are outside the expression of $M_{i1}$ have their potential mediators under the same treatment status as the treatment indicator to transmit the treatment effect.  {\tt  GADE} essentially represents the portion of {\tt  ATE} that is not transmitted by the induced change in users' organic search clicks.  {\tt  GADE}, as a generalized direct effect, incorporates parts of effects of upstream and downstream mediators.  In most of our applications, the upstream and downstream mediators are unknown.  We use the two terms ``the direct effect'' and ``{\tt  GADE}'' interchangeably for it simplifies our discussion and does not cause any confusion.

In $\text{{\tt  GACME}}(1)$, to extract the effect of the induced change in the users' clicks on the organic search results from {\tt  ATE}, the upstream mediators $\boldsymbol{M}_{i0}$ that are outside the expression of $M_{i1}$ are fixed at their potential mediators $\boldsymbol{M}_{i0}(1)$.  Only the upstream mediators $\boldsymbol{M}_{i0}$ that are inside the expression of $M_{i1}$ have their potential mediators under the same treatment status as the potential mediator of $M_{i1}$ to transmit the effect of the mediator $M_{i1}$.  Hence, {\tt  GACME} essentially represents the portion of {\tt  ATE} that is transmitted by the induced change in users' clicks on the organic search results.

In Figure~\ref{fig: mediation2}, {\tt  GADE} captures the causal effect of the treatment $T_i$ that goes through all the channels that do not have $M_{i1}$: $T\rightarrow Y$, $T\rightarrow \boldsymbol{M}_0\rightarrow Y$, $T\rightarrow \boldsymbol{M}_0\rightarrow \boldsymbol{M}_2\rightarrow Y$, $T\rightarrow \boldsymbol{M}_2\rightarrow Y$; {\tt  GACME} captures the causal effect of the treatment $T_i$ that goes through all the channels that have $M_{i1}$: $T\rightarrow M_1\rightarrow Y$, $T\rightarrow \boldsymbol{M}_0\rightarrow M_1\rightarrow Y$, $T\rightarrow M_1\rightarrow \boldsymbol{M}_2\rightarrow Y$, and $T\rightarrow \boldsymbol{M}_0\rightarrow M_1\rightarrow \boldsymbol{M}_2\rightarrow Y$.

\subsection{Linear Structural Equation Model}
Assuming linear relationships among all variables, measured and unmeasured, we use {\tt  LSEM} to parametrize their causal relationships.~\footnote{In this paper, we follow the literature that does not consider the interaction between mediators~\citep{Pearl2014InterpretationMediation.}.  Yet, it is a good topic for future research.}
\begin{definition}[{\tt  LSEM} with Multiple Causally-Dependent Mediators]
\begin{align}
\boldsymbol{M}_{i0} ={}& \boldsymbol{\alpha}_0 + \boldsymbol{\beta}_0T_i + \boldsymbol{e}_{i0}\label{um01}
\\
M_{i1} ={}& \alpha_1 + \beta_1T_i + \boldsymbol{\psi}_1^\intercal \boldsymbol{M}_{i0} + \boldsymbol{\xi}_1^\intercal \boldsymbol{M}_{i0}T_i + e_{i1}\label{um02}
\\
\begin{split}\label{um03}
\boldsymbol{M}_{i2} ={}& \boldsymbol{\alpha}_2 + \boldsymbol{\beta}_2T_i + \boldsymbol{\Psi}_2\boldsymbol{M}_{i0} + \boldsymbol{\psi}_3 M_{i1}\\
         &+ \boldsymbol{\Xi}_2 \boldsymbol{M}_{i0}T_i +                               \boldsymbol{\xi}_3 M_{i1}T_i + \boldsymbol{e}_{i2}
\end{split}
\\
\begin{split}\label{um04}
Y_i ={}& \alpha_3 + \beta_3T_i + \boldsymbol{\gamma}_0^\intercal \boldsymbol{M}_{i0} + \gamma_1 M_{i1} + \boldsymbol{\gamma}_2^\intercal \boldsymbol{M}_{i2}\\
       &+ \boldsymbol{\kappa}_0^\intercal \boldsymbol{M}_{i0}T_i + \kappa_1 M_{i1}T_i + \boldsymbol{\kappa}_2^\intercal \boldsymbol{M}_{i2}T_i + e_{i3}
\end{split}
\shortintertext{where}
\boldsymbol{M}_{i0} :={}& (M_{i01}, \cdots, M_{i0k}, \cdots, M_{i0K})^\intercal\nonumber
\\
\boldsymbol{M}_{i2} :={}& (M_{i21}, \cdots, M_{i2j}, \cdots, M_{i2J})^\intercal\nonumber
\\
\boldsymbol{e}_{i0} :={}& (e_{i01}, \cdots, e_{i0k}, \cdots, e_{i0K})^\intercal\nonumber
\\
\boldsymbol{e}_{i2} :={}& (e_{i21}, \cdots, e_{i2j}, \cdots, e_{i2J})^\intercal\nonumber
\end{align}
$e_{i0k}$, $k=1,\cdots,K$, $e_{i1}$, $e_{i2j}$, $j=1,\cdots,J$, and $e_{i3}$ are normalized to be mean zeros for all potential errors\footnote{See, e.g., \citet{Imai2010IdentificationEffects, Imai2013IdentificationExperiments, Keele2015IdentifyingAnalysis} for the assumption of mean zeros of potential errors.}.
\end{definition}
Equations~\eqref{um01} and~\eqref{um03} can be viewed as the reduced forms of structural equations that parametrize causal relationships among multiple upstream mediators and among multiple downstream mediators.  For example, the structural equations of two causally-dependent upstream mediators can be
\begin{align}
M_{i01} ={}& \alpha_{01} + \beta_{01}T_i + e_{i01}\label{eq: m21}
\\
M_{i02} = {}& \alpha_{02} + \beta_{02}T_i + \psi M_{i01} + \xi M_{i01}T_i + e_{i02}.\label{eq: m22}
\end{align}
We can write out their reduced-forms, in which Equation~\eqref{eq: m21} does not change, while Equation~\eqref{eq: m22} becomes   
\begin{align*}
M_{i02} ={}& \alpha_{02}' + \beta_{02}'T_i + e_{i02}'
\end{align*}
where $\alpha_{02}'=\alpha_{02} + \psi\alpha_{01}$, $\beta_{02}'=\beta_{02} + \psi\beta_{01} + \xi\alpha_{01} + \xi\beta_{01}$, and $e_{i02}' = (\psi + \xi T_i)e_{i01} + e_{i02}$.
\subsection{Sequential Ignorability}\label{sec:si_multiplemediators}
We generalize the {\tt SI} of \citet{Imai2010IdentificationEffects} by explicitly incorporating multiple causally-dependent mediators.

{\tt SI} is an unverifiable assumption in {\tt CMA}.  There are many in-depth discussions around {\tt SI} (see, e.g., the discussion between \citet{Pearl2014InterpretationMediation.,Pearl2014ReplyAnalysis} and \citet{Imai2014CommentAnalysis.}).  Researchers generally agree that restrictions imposed by {\tt SI} ``play a role in observational studies but not in studies where treatment is randomized.''~\citep{Pearl2014ReplyAnalysis}  {\tt SI} is the natural extension of strong ignorability~\citep{Rosenbaum1983TheEffects}.  {\tt SI} pictures sequential treatment assignments (ideal interventions) in randomized experiments.  First of all, the treatment is independent of all potential outcomes and potential mediators (i.e., ignorable) and its probability is strictly between 0 and 1, which is exactly the same as strong ignorability and guaranteed by random assignment.  Then, for each mediator, conditional on the treatment and its upstream mediators, each of its potential mediators behave like the treatment and are ignorable to the potential outcomes and the potential mediators of its downstream mediators.  
\begin{assumption}[{\tt SI} with Multiple Causally-Dependent Mediators]\label{assumption: SIMM}
\begin{align}
&\left\{
    \begin{array}{l}
    Y_i(t''', \boldsymbol{m}_0'', m_1', \boldsymbol{m}_2)\\
    M_{i2j}(t'', \boldsymbol{m}_0', m_1)\\
    M_{i1}(t', \boldsymbol{m}_0)\\
    M_{i0k}(t)
    \end{array}
\right\} 
\independent T_i\label{eq: multipleSI1}
\\
&\left\{
    \begin{array}{l}
    Y_i(t''', \boldsymbol{m}_0'', m_1', \boldsymbol{m}_2)\\
    M_{i2j}(t'', \boldsymbol{m}_0', m_1)\\
    M_{i1}(t', \boldsymbol{m}_0)\\
    \end{array}
\right\}
\independent M_{i0k}(t) \vert T_i=t\label{eq: multipleSI2}
\\
&\left\{
    \begin{array}{l}
    Y_i(t'', \boldsymbol{m}_0'', m_1', \boldsymbol{m}_2)\\
    M_{i2j}(t', \boldsymbol{m}_0', m_1)\\
    \end{array}
\right\}
\independent M_{i1}(t, \boldsymbol{m}_0) 
\Big\vert 
    \begin{array}{l}
    T_i=t\\
    \boldsymbol{M}_{i0}=\boldsymbol{m}_0
    \end{array}\label{eq: multipleSI3}
\\
&Y_i(t'', \boldsymbol{m}_0', m_1', \boldsymbol{m}_2)
\independent M_{i2j}(t, \boldsymbol{m}_0, m_1)
\Bigg\vert 
    \begin{array}{l}
    T_i=t\\
    \boldsymbol{M}_{i0}=\boldsymbol{m}_0\\
    M_{i1}=m_1
    \end{array}\label{eq: multipleSI4}
\\
&0<\mathbb{P}(T_i=t)<1 \label{eq: multipleSI5}
\\
&0<\mathbb{P}(M_{i0k}(t)=m_{0k} \vert  T_i=t)<1 \label{eq: multipleSI6}
\\
&0<\mathbb{P}(M_{i1}(t, \boldsymbol{m}_0)=m_{1} \vert  T_i=t,\boldsymbol{M}_{i0}=\boldsymbol{m}_0)<1 \label{eq: multipleSI7}
\\
&0<\mathbb{P}
\Bigg(
M_{i2j}(t, \boldsymbol{m}_0, m_1)=m_{2j}
\Bigg\vert 
    \begin{array}{l}
    T_i=t\\
    \boldsymbol{M}_{i0}=\boldsymbol{m}_0\\
    M_{i1}=m_1
    \end{array}
    \Bigg)
    <1 \label{eq: multipleSI8}
\intertext{where}
&\boldsymbol{m}_{0} := (m_{01}, \cdots, m_{0k}, \cdots, m_{0K})^\intercal\nonumber\\
&\boldsymbol{m}_{0}' := (m_{01}', \cdots, m_{0k}', \cdots, m_{0K}')^\intercal\nonumber\\
&\boldsymbol{m}_{0}'' := (m_{01}'', \cdots, m_{0k}'', \cdots, m_{0K}'')^\intercal\nonumber\\
&\boldsymbol{m}_{2} := (m_{21}, \cdots, m_{2j}, \cdots, m_{2J})^\intercal\nonumber
\end{align}
$\forall t, t', t'', t'''\in\left\lbrace 0,1\right\rbrace$, $\forall m_{0k}, m_{0k}', m_{0k}''\in\mathcal{M}_{0k}$, $k=1,\cdots,K$, $\forall m_1,m_1'\in\mathcal{M}_1$, and $\forall m_{2j}\in\mathcal{M}_{2j}$, $j=1,\cdots,J$.
\end{assumption}
In the literature, the violation of {\tt SI} in randomized experiments is typically due to covariates that confound the mediator-outcome relationship, which can be either pre-treatment or post-treatment covariates (i.e., causally-dependent mediators).  In Etsy's A/B tests, because of our ``guest checkout'' function (i.e., users can make purchases without registering at Etsy) and increasingly stringent privacy protection, it is difficult to obtain users' pre-treatment covariates, such as age, gender, and education level.  \citet{Imai2010IdentificationEffects} offers a sensitivity analysis to assess impacts of potential unmeasured pre-treatment confounders upon {\tt  ADE} and {\tt  ACME} under their {\tt SI}.  To explore a sensitivity analysis under our generalized {\tt SI} is a good topic for future research.  In addition, because we model all post-treatment variables, including measured and unmeasured causally-dependent mediators, in the generalized {\tt SI}, post-treatment confounders are no longer threats to our generalized {\tt SI}.

In the literature of multiple causally-dependent mediators, \citet{Imai2013IdentificationExperiments} and \citet{Robins2003SemanticsEffects} use a relaxed {\tt SI} (Assumption 2 and the {\tt FFRCISTG} respectively) with ``no interaction between treatment and mediators'' and ``no unmeasured mediators'' assumptions to nonparametrically identify {\tt  ACME}.  For our practical purposes, we allow \textit{the interaction between treatment and mediators}, and \textit{unmeasured mediators}, but assume {\tt  LSEM} and the generalized {\tt SI}.

\subsection{Identification}
With {\tt  LSEM}, {\tt SI}, and our definitions of {\tt  GACME} and {\tt  GADE}, we are ready to present important results about the identification.
\begin{lemma}[Properties of {\tt  LSEM} under Assumption~\ref{assumption: SIMM}]\label{lemma:lsem}
Consider {\tt  LSEM} defined in  Equations~\eqref{um01},~\eqref{um02},~\eqref{um03}, and~\eqref{um04} with all potential errors being mean zeros.  Under Assumption~\ref{assumption: SIMM},
\begin{align*}
\mathbb{E}(\boldsymbol{e}_{i0}\vert  T_i)&=\boldsymbol{0}
\\
\mathbb{E}(e_{i1}\vert  T_i, \boldsymbol{M}_{i0})&=0
\\
\mathbb{E}(\boldsymbol{e}_{i2}\vert  T_i, \boldsymbol{M}_{i0}, M_{i1})&=\boldsymbol{0}
\\
\mathbb{E}(e_{i3}\vert  T_i, \boldsymbol{M}_{i0}, M_{i1}, \boldsymbol{M}_{i2})&=0.
\end{align*}
\end{lemma}
The proof is in Appendix~\ref{appendix: lemma_lsem}.  Lemma~\ref{lemma:lsem} is an extension of Theorem~2 of \citet{Imai2010IdentificationEffects} who discuss the case of only one mediator.

\begin{definition}[Linear Regression System]
Define a linear regression system as
\begin{align}
M_{i1} &= \theta_{M_10} + \theta_{M_11}T_i + \mu_{M_1}\label{equation: lrs1}\\
Y_i    &= \theta_{Y0}   + \theta_{Y1}T_i   + \theta_{Y2}M_{i1} + \theta_{Y3}M_{i1}T_i + \mu_Y\label{equation: lrs2}.
\end{align}
\end{definition}
The linear regression system is constructed using all measured variables (the outcome $Y_i$, the treatment $T_i$, and the mediator $M_{i1}$).  Different from {\tt  LSEM}, it does not describe any causal relationships. Its coefficients are not necessarily correspond to the coefficients in {\tt  LSEM}.  Note that, we do not impose any assumption on its error terms.  Our assumptions are all in {\tt  LSEM} and Assumption~\ref{assumption: SIMM}.  

Lemma~\ref{lemma:lsem} implies the identification of coefficients in the linear regression system (Lemma~\ref{lemma:lrs}). 
\begin{lemma}[Identification of Coefficients in Linear Regression System]\label{lemma:lrs}
Consider {\tt  LSEM} defined in  Equations~\eqref{um01},~\eqref{um02},~\eqref{um03}, and~\eqref{um04} with all potential errors being mean zeros.  Under Assumption~\ref{assumption: SIMM}, when only $Y_i$, $T_i$, and $M_{i1}$ are measured, all the coefficients in Equations~\eqref{equation: lrs1} and~\eqref{equation: lrs2} are identified.
\end{lemma}
The proof is in Appendix~\ref{appendix: lemma_lrs}.  It uses \textit{strict exogeneity} conditions derived from Lemma~\ref{lemma:lsem}.  

Finally, we show that {\tt  GADE} and {\tt  GACME} can be obtained from the identified coefficients of the linear regression system.
\begin{theorem}[Identification of {\tt  GACME} and {\tt  GADE} with Multiple Unmeasured Causally-Dependent Mediators] \label{theorem: identification}
Consider {\tt  LSEM} defined in  Equations~\eqref{um01},~\eqref{um02},~\eqref{um03}, and~\eqref{um04} with all potential errors being mean zeros.  Under Assumption~\ref{assumption: SIMM}, when only $Y_i$, $T_i$, and $M_{i1}$ are measured, {\tt  GADE} and {\tt  GACME} are identified, and
\begin{align*}
\text{{\tt  GADE}}(t)\ &= \theta_{Y1} + \theta_{Y3}(\theta_{M_10} + \theta_{M_11}t)\\
\text{{\tt  GACME}}(t)\ &= \theta_{M_11}(\theta_{Y2} + \theta_{Y3}t)
\end{align*}
for $t=0,1$, where $\theta_{M_10}$, $\theta_{M_11}$, $\theta_{Y1}$, $\theta_{Y2}$, and $\theta_{Y3}$ are the coefficients in Equations~\eqref{equation: lrs1} and~\eqref{equation: lrs2}.
\end{theorem}
The proof is in Appendix~\ref{appendix: theorem_1}.

\section{Estimation and Hypothesis Testing}\label{sec:estimation}
Theorem~\ref{theorem: identification} shows {\tt  GACME} and {\tt  GADE} can be estimated via two linear regression equations, which are easy to estimate in practice.

In Theorem~\ref{theorem: identification}, {\tt  GACME} and {\tt  GADE} are continuous functions of coefficients of Equations~\eqref{equation: lrs1} and~\eqref{equation: lrs2}.  Based on continuous mapping theorem, if estimators of regression coefficients are consistent to their true values, the estimators of {\tt  GACME} and {\tt  GADE} that are functions of those consistent estimators of regression coefficients will be consistent to the true values of {\tt  GACME} and {\tt  GADE}.  Therefore, the estimation of {\tt  GACME} and {\tt  GADE} consists of two steps: (i) estimating the regression coefficients of Equations~\eqref{equation: lrs1} and~\eqref{equation: lrs2}, and (ii) calculating the estimates of {\tt  GADE} and {\tt  GACME} by their closed-form expressions in Theorem~\ref{theorem: identification} and the estimates of regression coefficients from Step~(i).  The hypothesis testing of {\tt  GACME} and {\tt  GADE} is based on central limit theorem and Delta method.

\subsection{Iterative General Method of Moments}
In the proof of Lemma~\ref{lemma:lrs} (Appendix~\ref{appendix: lemma_lrs}), the conditional mean zero conditions from Lemma~\ref{lemma:lsem} imply \textit{strict exogeneity} conditions that guarantees that the regression coefficients in Equations~\eqref{equation: lrs1} and~\eqref{equation: lrs2} can be consistently and separately estimated via \textit{ordinary least squares}~\citep{Wooldridge2010EconometricData}.  However, the standard errors of {\tt  GADE} and {\tt  GACME} depend on variances of coefficient estimators of the two equations.  Moreover, the error terms $\mu_{M_1}$ and $\mu_Y$ of the two equations are correlated if unmeasured causally-dependent mediators exist.  Therefore, it is more efficient to estimate the two equations jointly than separately.

We suggest using iterative general method of moments ({\tt  ITGMM})~\citep{Hansen1982LargeEstimators,Hansen1996Finite-sampleEstimators} with the heteroskedasticity and autocorrelation consistent (HAC) covariance matrix~\citep{Newey1987AMatrix,Andrews1991HeteroscedasticityEstimation} to jointly estimate Equations~\eqref{equation: lrs1} and~\eqref{equation: lrs2}.  The {\tt  ITGMM} estimators are consistent to true values, and are efficient when heteroskedasticity or autocorrelation exists.

The complete algorithm is in Appendix~\ref{appendix: itgmm}.  An open sourced \verb|R| package \verb|gmm| from~\citet{Chausse2010ComputingR} allows us to implement {\tt  ITGMM} conveniently.

\subsection{Hypothesis Testing}
To conduct hypothesis tests on $H_0$: $\text{{\tt  GADE}} = 0$ and $H_0$: $\text{{\tt  GACME}} = 0$, we suggest using central limit theorem and Delta method to construct test statistics that asymptotically follow standard normal distribution.

According to central limit theorem, the estimators $\widehat{\text{{\tt  GADE}}}$ and $\widehat{\text{{\tt  GACME}}}$ will converge in distribution to normal distributions that have true values of these causal estimands as means and asymptotic variances (i.e., $\text{Avar}(\widehat{\text{{\tt  GADE}}})$ and $\text{Avar}(\widehat{\text{{\tt  GACME}}})$).  Hence, we need to estimate $\text{Avar}(\widehat{\text{{\tt  GADE}}})$ and $\text{Avar}(\widehat{\text{{\tt  GACME}}})$.

Because $\text{{\tt  GADE}}$ and $\text{{\tt  GACME}}$ have continuous first partial derivatives with respect to $\theta$, we employ Delta method to express $\text{Avar}(\widehat{\text{{\tt  GADE}}})$ and $\text{Avar}(\widehat{\text{{\tt  GACME}}})$ as functions of $\theta$ and asymptotic variances and covariances of $\hat{\theta}$.  By replacing all unknown quantities in the expressions with their consistent estimators, according to continuous mapping theorem, we can get consistent estimators of $\text{Avar}(\widehat{\text{{\tt  GADE}}})$ and $\text{Avar}(\widehat{\text{{\tt  GACME}}})$ (i.e., $\widehat{\text{Avar}}(\widehat{\text{{\tt  GADE}}}(t))$ and $\widehat{\text{Avar}}(\widehat{\text{{\tt  GACME}}}(t))$).  The derived formulas are in Appendix~\ref{appendix: delta_method}.

The test statistics under $H_0$: $\text{{\tt  GADE}} = 0$ and $H_0$: $\text{{\tt  GACME}} = 0$ are:
\begin{align*}
    \frac{\widehat{\text{{\tt  GADE}}}(t)}{\widehat{\text{Avar}}(\widehat{\text{{\tt  GADE}}}(t))} \xrightarrow{d} N(0, 1)\\
    \frac{\widehat{\text{{\tt  GACME}}}(t)}{\widehat{\text{Avar}}(\widehat{\text{{\tt  GACME}}}(t))} \xrightarrow{d} N(0, 1),
\end{align*}
respectively.
\section{Applications at Etsy}\label{sec:experiments}
We present two real cases to show how we use the analysis for Etsy's A/B tests to derive insights on our products.
\subsection{Recommendation Module A/B Test}
\begin{table*}
  \caption{Recommendation Module A/B Test}
  \label{tab:lpssv2_AB}
\begin{adjustbox}{max width=\textwidth}
\begin{threeparttable}
\begin{tabular}{lccccc}
\hline
 &  & \multicolumn{4}{c}{Two-Tailed $p$-Value, $H_0$: treatment = control} \\ Outcome
 & \% Change = {\tt  ATE}/Mean of Control & z-Test & MW U-Test & Prop. z-Test & Fisher's Exact Test \\ \hline
Recommendation Clicks & 28.3131\% & 0.0000 & 0.0000 &  &  \\
Conversion & 0.2202\% &  &  & 0.3638 & 0.3634 \\
{\tt GMV} & -0.2518\% & 0.7014 & 0.4023 &  &  \\
Organic Search Clicks & -1.3658\% & 0.0000 & 0.0000 &  &  \\ \hline
\end{tabular}
    \begin{tablenotes}
      \small
          \item 1) MW U-test is Mann-Whitney U-test.  Prop. z-test is two-proportion z-test.
          \item 2) For z-test, prop. z-test, and Fisher's exact test, $H_0$: the treatment and the control have equal means of the metric.  For MW U-test, $H_0$: the treatment and the control have equal distributions of the metric.
          \item 3) All the distributions of non-binary metrics are highly skewed.  Instead of t-test, z-test is used for the null hypothesis of mean equivalence based on asymptotical normality.
    \end{tablenotes}
\end{threeparttable}
\end{adjustbox}
\end{table*}
\begin{table}
  \caption{Estimates of Causal Effects for Recommendation Module A/B Test, Mediator is Organic Search Clicks}
  \label{tab:lpssv2_mediation}
\begin{adjustbox}{max width=\textwidth}
\begin{threeparttable}
\begin{tabular}{lcccc}
\hline
 & \multicolumn{2}{c}{Outcome: Conversion} & \multicolumn{2}{c}{Outcome: {\tt GMV}} \\
Effect & \% Change & Std Error & \% Change & Std Error \\ \hline
$\text{{\tt  GADE}}(0)$ & 0.4959\%* & 0.000272 & 0.1681\% & 0.037515 \\
$\text{{\tt  GADE}}(1)$ & 0.4905\%* & 0.000271 & 0.1700\% & 0.037294 \\
$\text{{\tt  GACME}}(0)$ & -0.2703\%*** & 0.000047 & -0.4219\%*** & 0.003701 \\
$\text{{\tt  GACME}}(1)$ & -0.2757\%*** & 0.000049 & -0.4200\%*** & 0.003733 \\
{\tt  ATE} & 0.2202\% & 0.000275 & -0.2518\% & 0.037582 \\ \hline
\end{tabular}
    \begin{tablenotes}
      \small
      \item 1) \% Change = Effect/Mean of Control
      \item 2) `***' $p<0.001$, `**' $p<0.01$, `*' $p<0.05$, `.' $p<0.1$.  Two-tailed $p$-value is derived from z-test for $H_0$: the effect is zero, which is based on asymptotical normality.
    \end{tablenotes}
\end{threeparttable}
\end{adjustbox}
\end{table}
\begin{table*}
  \caption{Promoted Listing A/B Test}
  \label{tab:prolist50_AB}
\begin{adjustbox}{max width=\textwidth}
\begin{threeparttable}
\begin{tabular}{lccccc}
\hline
 &  & \multicolumn{4}{c}{Two-Tailed $p$-Value, $H_0$: treatment = control} \\ Outcome
 & \% Change = {\tt  ATE}/Mean of Control & z-Test & MW U-Test & Prop. z-Test & Fisher's Exact Test \\ \hline
Promoted Listings &  &  &  &  &  \\
\multicolumn{1}{r}{Revenue} & 2.7044\% & 0.0000 & 0.0000 &  &  \\
\multicolumn{1}{r}{Clicks} & 8.9927\% & 0.0000 & 0.0000 &  &  \\
\multicolumn{1}{r}{Click Through} & 5.8722\% &  &  & 0.0000 & 0.0000 \\
Conversion & -0.3709\% &  &  & 0.3379 & 0.3379 \\
{\tt GMV} & -1.3456\% & 0.1568 & 0.3151 &  &  \\
Organic Search Clicks & -0.8106\% & 0.0004 & 0.0082 &  &  \\ \hline
\end{tabular}
    \begin{tablenotes}
      \small
          \item 1) MW U-test is Mann-Whitney U-test.  Prop. z-test is two-proportion z-test.
          \item 2) For z-test, prop. z-test, and Fisher's exact test, $H_0$: the treatment and the control have equal means of the metric.  For MW U-test, $H_0$: the treatment and the control have equal distributions of the metric.
          \item 3) All the distributions of non-binary metrics are highly skewed.  Instead of t-test, z-test is used for the null hypothesis of mean equivalence based on asymptotical normality.
    \end{tablenotes}
\end{threeparttable}
\end{adjustbox}
\end{table*}
\begin{table}
  \caption{Estimates of Causal Effects for Promoted Listing A/B Test, Mediator is Organic Search Clicks}
  \label{tab:prolist50_mediation}
\begin{adjustbox}{max width=\textwidth}
\begin{threeparttable}
\begin{tabular}{lcccc}
\hline
 & \multicolumn{2}{c}{Outcome: Conversion} & \multicolumn{2}{c}{Outcome: {\tt GMV}} \\
Effect & \% Change & Std Error & \% Change & Std Error \\ \hline
$\text{{\tt  GADE}}(0)$ & -0.1448\% & 0.000203 & -1.0682\% & 0.022971 \\
$\text{{\tt  GADE}}(1)$ & -0.1472\% & 0.000202 & -1.0526\% & 0.022868 \\
$\text{{\tt  GACME}}(0)$ & -0.2237\%*** & 0.000034 & -0.2930\%*** & 0.002030 \\
$\text{{\tt  GACME}}(1)$ & -0.2261\%*** & 0.000034 & -0.2774\%*** & 0.001924 \\
{\tt  ATE} & -0.3709\% & 0.000205 & -1.3456\% & 0.023033 \\ \hline
\end{tabular}
    \begin{tablenotes}
      \small
      \item 1) \% Change = Effect/Mean of Control
      \item 2) `***' $p<0.001$, `**' $p<0.01$, `*' $p<0.05$, `.' $p<0.1$.  Two-tailed $p$-value is derived from z-test for $H_0$: the effect is zero, which is based on asymptotical normality.
    \end{tablenotes}
\end{threeparttable}
\end{adjustbox}
\end{table}

A typical A/B test of a recommendation module at Etsy commonly involves the change in the recommendation algorithm or its user interface or sometimes both. Product owners seek increases in key business metrics, such as conversation rate or \textit{Gross Merchandise Value} ({\tt GMV}), from such a change.  Note that, the key business metrics are usually computed at the sitewide level because the ultimate goal of the improvement of any product on the platform is to make better user experience about the whole site.

 We take one such experiment as an example and show its main results in Table~\ref{tab:lpssv2_AB}. We can see that the change in the recommendation module results in a significant $28.3131\%$ increase in users' clicks on the recommendation module, but an insignificant $0.2202\%$ increase in conversion rate and an insignificant $0.2518\%$ decrease in {\tt GMV}.  We wavered over whether to ramp down the change because there are no significant gains for the whole marketplace concerning the overall conversion rate and {\tt GMV} while the user engagement in the recommendation module indeed increases.  Note that, the change also significantly reduces users' clicks on organic search results by $1.3658\%$.  We conjectured that, the improved recommendation module successfully suggests listings that satisfy users' needs so that users do not need the organic search as much as usual, and the negative effects of the induced reduction of the user engagement in the organic search can offset, or even exceed, the positive effects of the improvement of the recommendation module, which leads to an insignificant increase in conversion rate and an insignificant decrease in {\tt GMV}.  If the conjecture is true, then we should ramp up the new version of the recommendation module and improve the organic search in the future so that they can jointly work better later.  The conjecture needs {\tt CMA} to justify.  The key task is to distinguish the direct effects of the change in the recommendation module and the effects of the induced reduction of user engagement in the organic search (i.e., the indirect effects of the change in the recommendation module).

Table~\ref{tab:lpssv2_mediation} shows the results from {\tt CMA}.  We firstly discuss the results where conversion rate is the outcome.  On average, holding a user's clicks on organic search results as if she had been presented with the old version of the recommendation module, the change in the recommendation module significantly increases her conversion rate by $0.4959\%$.  So the average direct effect in the control group (i.e., $\text{{\tt  GADE}}(0)$) is $0.4959\%$.  On average, if a user was presented with the new version of the recommendation module, the induced reduction of her clicks on organic search results significantly decreases her conversion rate by $0.2757\%$.  So the average indirect effect in the treatment group (i.e., $\text{{\tt  GACME}}(1)$) is $-0.2757\%$.  The indirect effect partially offsets the direct effect, which makes the total effect {\tt  ATE} (see Equation~\ref{eq: DecomposeATE}) less than half of the direct effect.  In addition, {\tt  ATE}, as a summation of the two effects, incorporates sampling errors of the two effects, which makes its standard error greater than those of the two effects.  {\tt  ATE}'s small magnitude and large standard error lead to its insignificance.

Note that, the direct effect and the indirect effect vary slightly between treatment and control groups.  On average, holding a user's clicks on organic search results as if she had been presented with the new version of the recommendation module, the change in the recommendation module significantly increases her conversion rate by $0.4905\%$ (i.e., $\text{{\tt  GADE}}(1)$).  On average, if a user was presented with the old version of the recommendation module, the induced reduction of her clicks on organic search results significantly decreases her conversion rate by $0.2703\%$ (i.e., $\text{{\tt  GACME}}(0)$).  As before, the summation of these two effects (see Equation~\ref{eq: DecomposeATE}) is {\tt  ATE}.

Next, we move to the results where outcome is {\tt GMV}.  The direct effects of the change in the recommendation module (i.e., $\text{{\tt  GADE}}(0)$ and $\text{{\tt  GADE}}(1)$) are positive but insignificant.  The indirect effects, the effects of the induced reduction of users' clicks on organic search results (i.e., $\text{{\tt  GACME}}(0)$ and $\text{{\tt  GACME}}(1)$), are significantly negative. The negative indirect effects exceed the positive direct effects in magnitude, which makes the total effect {\tt  ATE} negative.

To sum up, the change in the recommendation module should have brought us more significant increases of conversion rate than what the A/B test shows.  Its positive contribution is partially offset by the negative impact from the induced reduction of user engagement in the organic search.  The impaired organic search cannot be ignored because the effects from the induced reduction of users' clicks on organic search results are significantly negative regarding the conversion and {\tt GMV}.  The analysis justifies our conjecture on the underlying causal mechanism.  We launched the new version of the recommendation module and decided to improve the organic search to better work with the recommendation module in our next step of product development.
\subsection{Promoted Listing A/B Test}
Table~\ref{tab:prolist50_AB} shows the results from an A/B test of promoted listings in Etsy.  We can see the change in promoted listings significantly lifts key metrics of the product, which include revenue (by $2.7044\%$), the number of clicks (by $8.9927\%$), and {\tt CTR} (by $5.8722\%$).  From the perspective of the advertising service, the change is successful.  It is noteworthy that, however, the change significantly reduces users' clicks on organic search results by $0.8106\%$.  A reasonable conjecture is that the improved promoted listings successfully match users' intention and divert users' attention from organic search results on same search engine result pages where promoted listings and organic search results are presented together. Also, note that, conversion rate and {\tt GMV} have insignificant decreases. We further conjecture that, the induced reduction of user engagement in the organic search has negative effects upon conversion rate and {\tt GMV}, which are partially offset by the positive effects of the change in promoted listings because promoted listings are also channels for users to explore Etsy's marketplace.  Again, these conjectures need {\tt CMA} to justify.  The key task is to distinguish the direct effects of the change in promoted listings and the effects of the induced reduction of user engagement in the organic search (i.e., the indirect effects of the change in promoted listings).

Table~\ref{tab:prolist50_mediation} shows results from {\tt CMA}.  The direct effects of the change in promoted listings (i.e., $\text{{\tt  GADE}}(0)$ and $\text{{\tt  GADE}}(1)$) upon conversion rate and {\tt GMV} are insignificantly negative.  However, the indirect effects, which are the effects of the induced reduction of users' clicks on organic search results (i.e., $\text{{\tt  GACME}}(0)$ and $\text{{\tt  GACME}}(1)$), upon conversion rate and {\tt GMV} are significantly negative.  For conversion, the indirect effects are greater than the direct effects in magnitude. For {\tt GMV}, the opposite is true.  {\tt  ATE}, as a summation of the direct effect and the indirect effect (see Equation~\ref{eq: DecomposeATE}), is negative and has a greater magnitude than the two effects.  Again, {\tt  ATE} incorporates sampling errors of the two effects, which makes its standard error greater than those of the two effects.

To sum up, the change in promoted listings, which improves advertising, hurts the marketplace in terms of both conversion rate and {\tt GMV} though the impacts are statistically insignificant.  The impaired organic search cannot be ignored because the impacts from the induced reduction of user engagement in the organic search are significantly negative regarding the conversion rate ($-0.2237\%$ and $-0.2261\%$) and {\tt GMV} ($-0.2930\%$ and $-0.2774\%$).

\section{Conclusion}\label{sec:conclusion}
The A/B test plays a crucial role in evaluating product changes in internet companies.  Decisions on the product change are typically based on the effects of the change in the tested product upon the key business metrics, such as conversion rate or {\tt GMV}.  Many A/B tests show that the change in the tested product can induce users to change their behaviors in other products.  The effects that are calculated directly from an A/B test are essentially total effects, which are combinations of direct effects and indirect effects of the change in the tested product.  The indirect effects are the effects from the induced change in user behaviors in other products.  To better understand the change in the tested product, it is desirable to assess its direct effects on sitewide business metrics.  To optimize the whole website, it is not advisable to ignore the sitewide impacts from the induced changes in user behaviors in other products.  Unfortunately, neither direct effects nor indirect effects can be calculated directly from A/B tests.

We introduce {\tt CMA}, a formal statistical framework, to estimate direct and indirect effects of the change in the tested product upon sitewide business metrics by utilizing data from A/B tests.  However, an A/B test can have numerous unmeasured causally-dependent mediators, which break the assumptions of {\tt CMA} and invalidate the identification of direct and indirect effects.  To solve the problem, we generalize definitions of causal estimands and assumptions in the literature of {\tt CMA} by explicitly incorporating unmeasured mediators.  We show that the generalized direct and indirect effects are still useful for our evaluation purposes.  We prove that under the generalized assumptions, the two effects are identifiable and can be estimated through two linear regression equations.  We suggest using {\tt  ITGMM} and Delta method, which are well-developed methods in the literature of statistics and econometrics, for estimation and statistical inference.  The \verb|R| package \verb|gmm| allows us to conduct the estimation conveniently.

We apply our method to analyze two A/B tests of Etsy.  Through the analysis, we derive the estimates of direct and indirect effects of the changes in the tested products upon conversion and {\tt GMV}.  These estimates reveal the underlying causal mechanisms from the changes in the tested products to Etsy's marketplace and uncover the complex causal relationships between different products, all of which cannot be directly obtained from A/B tests.

\bibliographystyle{ACM-Reference-Format}
\balance 
\bibliography{Mendeley_references}

\newpage

\appendix
\section{Proof of Equation~(\ref{eq: DecomposeATE})}\label{appendix: eq_proof}
\begin{proof}
\begin{align*}
\text{{\tt  ATE}} = \mathbb{E}[Y_i(1,&\boldsymbol{M}_{i0}(1), M_{i1}(1, \boldsymbol{M}_{i0}(1)),\nonumber\\
&\boldsymbol{M}_{i2}(1, \boldsymbol{M}_{i0}(1), M_{i1}(1, \boldsymbol{M}_{i0}(1))))]\nonumber\\
-\mathbb{E}[Y_i(&0, \boldsymbol{M}_{i0}(0), M_{i1}(0, \boldsymbol{M}_{i0}(0)),\nonumber\\
&\boldsymbol{M}_{i2}(0, \boldsymbol{M}_{i0}(0), M_{i1}(0, \boldsymbol{M}_{i0}(0))))]\nonumber\\
= \text{{\tt  GADE}}&(0)+\text{{\tt  GACME}}(1) = \text{{\tt  GADE}}(1)+\text{{\tt  GACME}}(0).    
\end{align*}
\end{proof}

\section{The Algorithm of {\tt ITGMM}}\label{appendix: itgmm}
General method of moments (GMM) was developed as a generalization of the method of moments (a classical estimation method in statistics) by \citet{Hansen1982LargeEstimators} in 1982, who shared 2013 Nobel Prize in Economics in part for GMM.  To improve the finite sample performance of GMM, \citet{Hansen1996Finite-sampleEstimators} proposed two alternatives: {\tt ITGMM} and the continuous updated GMM (CUE).  CUE offers smaller second order asymptotic bias than {\tt ITGMM}, but requires numerical differentiation, which is computationally costly.  See \citet{Chausse2010ComputingR} for a detailed discussion.

Let $S_i$ denote the vector of $Y_i$, $T_i$, and $M_{i1}$ for user $i$.  Let $\theta$ denote the set of coefficients of the linear regression system of Equations~\eqref{equation: lrs1} and~\eqref{equation: lrs2} $\{$$\theta_{M_11}$, $\theta_{M_10}$, $\theta_{Y0}$, $\theta_{Y1}$, $\theta_{Y2}$, $\theta_{Y3}$$\}$.

Define functions:
\begin{align*}
g_{i1}(\theta, S_i) &= M_{i1} - \theta_{M_10} - \theta_{M_11}T_i\\
g_{i2}(\theta, S_i) &= T_i(M_{i1} - \theta_{M_10} - \theta_{M_11}T_i)\\
g_{i3}(\theta, S_i) &= Y_i - \theta_{Y0} - \theta_{Y1}T_i - \theta_{Y2}M_{i1} - \theta_{Y3}M_{i1}T_i\\
g_{i4}(\theta, S_i) &= T_i(Y_i - \theta_{Y0} - \theta_{Y1}T_i - \theta_{Y2}M_{i1} - \theta_{Y3}M_{i1}T_i)\\
g_{i5}(\theta, S_i) &= M_{i1}(Y_i - \theta_{Y0} - \theta_{Y1}T_i - \theta_{Y2}M_{i1} - \theta_{Y3}M_{i1}T_i)\\
g_{i6}(\theta, S_i) &= M_{i1}T_i(Y_i - \theta_{Y0} - \theta_{Y1}T_i - \theta_{Y2}M_{i1} - \theta_{Y3}M_{i1}T_i)
\end{align*}
Define $g_i(\theta, S_i)$ to be the vector obtained by stacking the above six values into one column vector.  The sample average of $g_i(\theta, S_i)$ is $\overbar{g}(\theta)=\frac{1}{N} \sum_{i=1}^N g_i(\theta, S_i)$ where $N$ is the number of users.  Define HAC matrix as $\hat{\Omega}(\theta) = \sum_{s=-(N-1)}^{N-1} k_h(s) \hat{\Gamma}_s(\theta)$
where $\hat{\Gamma}_s(\theta)=\frac{1}{N}\sum_{i=1}^N g_i(\theta, S_i)g_{i+s}(\theta, S_{i+s})^{\intercal}$, $k_h(s)$ is a kernel, and $h$ is the bandwidth that can be chosen using the procedures proposed by \citet{Newey1987AMatrix} and \citet{Andrews1991HeteroscedasticityEstimation}.  
\begin{algorithm}[h]
\caption{{\tt  ITGMM} from \citet{Chausse2010ComputingR}}\label{algorithm: itgmm}
\begin{algorithmic}[1]
\Input{$Y_i$, $T_i$, and $M_{i1}$, $i=1,\cdots,N$}
\Output $\hat{\theta}$

\State Compute the initial value $ \theta^{(0)}=\argmin_{\theta}\overbar{g}(\theta)^{\intercal}\overbar{g}(\theta)$

\State Compute the HAC matrix $\hat{\Omega}(\theta^{(0)})$\label{state2}

\State Compute $\theta^{(1)}=\argmin_{\theta}\overbar{g}(\theta)^{\intercal}[\hat{\Omega}(\theta^{(0)})]^{-1}\overbar{g}(\theta)$

\If{$\norm[\big]{\theta^{(0)} - \theta^{(1)}} < tol$} 
\State stops
\Else
\State $\theta^{(0)} \gets \theta^{(1)}$ and go to~\ref{state2}
\EndIf 

\State $\hat{\theta} \gets \theta^{(1)}$

\State \Return{$\hat{\theta}$}
\end{algorithmic}
\end{algorithm}

The {\tt  ITGMM} estimator $\hat{\theta}$ is approximately distributed as $N(\theta^{*}, \hat{V}/N)$ where $\hat{V} = {\frac{\partial \overbar{g}(\hat{\theta})}{\partial \theta}}^{\intercal} {\hat{\Omega}(\hat{\theta})}^{-1} {\frac{\partial \overbar{g}(\hat{\theta})}{\partial \theta}}$, $\hat{V}/N$ is the estimator of the asymptotic variance-covariance matrix of $\hat{\theta}$, and $\theta^{*}$ is the true value.

\section{Consistent Estimators of Asymptotic Variance of Estimators of Causal Effects}\label{appendix: delta_method}
The formulas are derived from Delta method.
\begin{align*}
\widehat{\text{Avar}}(\widehat{\text{{\tt  GADE}}}(t)) =&
\widehat{\text{Avar}}(\hat{\theta}_{Y1})
+ (\hat{\theta}_{M_10} + \hat{\theta}_{M_11}t)^2 \widehat{\text{Avar}}(\hat{\theta}_{Y3})
\\
&+ \hat{\theta}_{Y3}^2 \widehat{\text{Avar}}(\hat{\theta}_{M_10})
+ (\hat{\theta}_{Y3}t)^2 \widehat{\text{Avar}}(\hat{\theta}_{M_11})
\\
&+ 2\hat{\theta}_{Y3} \widehat{\text{Acov}}(\hat{\theta}_{Y1}, \hat{\theta}_{M_10})
\\
&+ 2(\hat{\theta}_{M_10}+\hat{\theta}_{M_11}t) \widehat{\text{Acov}}(\hat{\theta}_{Y1}, \hat{\theta}_{Y3})
\\
&+ 2(\hat{\theta}_{Y3}t) \widehat{\text{Acov}}(\hat{\theta}_{Y1}, \hat{\theta}_{M_11})
\\
&+ 2(\hat{\theta}_{M_10} + \hat{\theta}_{M_11}t)\hat{\theta}_{Y3} \widehat{\text{Acov}}(\hat{\theta}_{Y3}, \hat{\theta}_{M_10})
\\
&+ 2(\hat{\theta}_{M_10} + \hat{\theta}_{M_11}t)\hat{\theta}_{Y3}t \widehat{\text{Acov}}(\hat{\theta}_{Y3}, \hat{\theta}_{M_11})
\\
&+ 2\hat{\theta}_{Y3}^2t \widehat{\text{Acov}}(\hat{\theta}_{M_10}, \hat{\theta}_{M_11})\\
\widehat{\text{Avar}}(\widehat{\text{{\tt  GACME}}}(t)) =&
(\hat{\theta}_{Y2} + \hat{\theta}_{Y3}t)^2 \widehat{\text{Avar}}(\hat{\theta}_{M_11})
\\
&+ \hat{\theta}_{M_11}^2 \widehat{\text{Avar}}(\hat{\theta}_{Y2})
+ (\hat{\theta}_{M_11}t)^2 \widehat{\text{Avar}}(\hat{\theta}_{Y3})
\\
&+ 2(\hat{\theta}_{Y2}+\hat{\theta}_{Y3}t)\hat{\theta}_{M_11} \widehat{\text{Acov}}(\hat{\theta}_{M_11}, \hat{\theta}_{Y2})
\\
&+ 2(\hat{\theta}_{Y2}+\hat{\theta}_{Y3}t)\hat{\theta}_{M_11}t \widehat{\text{Acov}}(\hat{\theta}_{M_11}, \hat{\theta}_{Y3})
\\
&+ 2\hat{\theta}_{M_11}^2t \widehat{\text{Acov}}(\hat{\theta}_{Y2}, \hat{\theta}_{Y3})
\end{align*}
where $\hat{\theta}_{M_11}$, $\hat{\theta}_{M_10}$, $\hat{\theta}_{Y0}$, $\hat{\theta}_{Y1}$, $\hat{\theta}_{Y2}$, and $\hat{\theta}_{Y3}$ are {\tt  ITGMM} estimators, $\widehat{\text{Avar}}(\cdot)$ and $\widehat{\text{Acov}}(\cdot)$ are elements of $\hat{V}/N$ of {\tt  ITGMM}.

\section{Proof of Lemma~\ref{lemma:lsem}}\label{appendix: lemma_lsem}
\begin{proof}
Generally, the specification of unconditional mean zeros of potential errors in {\tt LSEM} does not imply conditional mean zeros of them.  However, Lemma~\ref{lemma:lsem} says, under our generalized {\tt SI}, it does.

Here we show the proof of $\mathbb{E}(\boldsymbol{e}_{i2}\vert T_i, \boldsymbol{M}_{i0}, M_{i1})=\boldsymbol{0}$.  Other conditional mean zero conditions can be easily derived using the same technique.
\begin{align*}
&\mathbb{E}(e_{i2j}(T_i, \boldsymbol{M}_{i0}, M_{i1})\vert T_i=t, \boldsymbol{M}_{i0}=\boldsymbol{m}_0, M_{i1}=m_1)\\
={}&\mathbb{E}(e_{i2j}(t, \boldsymbol{m}_0, m_1)\vert T_i=t, \boldsymbol{M}_{i0}=\boldsymbol{m}_0, M_{i1}=m_1)\\
={}&\mathbb{E}(e_{i2j}(t, \boldsymbol{m}_0, m_1)\vert T_i=t, \boldsymbol{M}_{i0}=\boldsymbol{m}_0)\\
={}&\mathbb{E}(e_{i2j}(t, \boldsymbol{m}_0, m_1)\vert T_i=t)\\
={}&\mathbb{E}(e_{i2j}(t, \boldsymbol{m}_0, m_1))\\
={}&0
\end{align*}
The first equality comes from the definition of potential outcomes. 
The second equality comes from the fact that
\begin{align*}
M_{i2j}(t, \boldsymbol{m}_0, m_1)\independent M_{i1}(t, \boldsymbol{m}_0)\vert T_i=t,\boldsymbol{M}_{i0}=\boldsymbol{m}_0
\end{align*} 
implies 
\begin{align*}
e_{i2j}(t, \boldsymbol{m}_0, m_1)\independent M_{i1}(t, \boldsymbol{m}_0)\vert T_i=t,\boldsymbol{M}_{i0}=\boldsymbol{m}_0 
\end{align*}
for all $t$, $\boldsymbol{m}_0$, and $m_1$.
The third equality comes from the fact that 
\begin{align*}
M_{i2j}(t, \boldsymbol{m}_0, m_1)\independent M_{i0k}(t) \vert T_i=t
\end{align*}
implies 
\begin{align*}
e_{i2j}(t, \boldsymbol{m}_0, m_1)\independent M_{i0k}(t) \vert T_i=t
\end{align*}
for all $t$, $\boldsymbol{m}_0$, and $m_1$.
The fourth equality comes from {\tt SI} that $T_i$ is independent of all potential outcomes.
The last equality comes from the fact that all potential outcomes of error terms of SEM can be normalized to be mean zeros.
For all $j$, $t$, $\boldsymbol{m}_0$, and $m_1$, we have 
\begin{align*}
\mathbb{E}(e_{i2j}(T_i, \boldsymbol{M}_{i0}, M_{i1})\vert T_i=t, \boldsymbol{M}_{i0}=\boldsymbol{m}_0, M_{i1}=m_1)=0
\end{align*}
so
\begin{align*}
\mathbb{E}(\boldsymbol{e}_{i2}\vert T_i, \boldsymbol{M}_{i0}, M_{i1})=\boldsymbol{0}.
\end{align*}
\end{proof}

\section{Proof of Lemma~\ref{lemma:lrs}}\label{appendix: lemma_lrs}
\begin{proof}
By Substituting Equation~\eqref{um01} into Equation~\eqref{um02}, Equations~\eqref{um01} and~\eqref{um03} into Equation~\eqref{um04}, we can get
\begin{align}
\begin{split}
M_{i1} ={}& (\alpha_1 + \boldsymbol{\psi}_1^\intercal\boldsymbol{\alpha}_0) \\
          &+ (\beta_1 + \boldsymbol{\psi}_1^\intercal\boldsymbol{\beta}_0 + \boldsymbol{\xi}_1^\intercal\boldsymbol{\alpha}_0 + \boldsymbol{\xi}_1^\intercal\boldsymbol{\beta}_0)T_i \\
          &+ (\boldsymbol{\psi}_1^\intercal + \boldsymbol{\xi}_1^\intercal  T_i)\boldsymbol{e}_{i0} + e_{i1} \label{eq: M1}
\end{split}
\\
\begin{split}
Y_i ={}& (\alpha_3 + \boldsymbol{\gamma}_0^\intercal\boldsymbol{\alpha}_0 + \boldsymbol{\gamma}_2^\intercal(\boldsymbol{\alpha}_2 + \boldsymbol{\Psi}_2\boldsymbol{\alpha}_0 + \boldsymbol{\Xi}_2\boldsymbol{\alpha}_0))\\
       &+ [\beta_3 + \boldsymbol{\gamma}_0^\intercal\boldsymbol{\beta}_0 + \boldsymbol{\gamma}_2^\intercal(\boldsymbol{\beta}_2 + \boldsymbol{\Psi}_2\boldsymbol{\beta}_0 + \boldsymbol{\Xi}_2\boldsymbol{\beta}_0)\\
       &+ \boldsymbol{\kappa}_0^\intercal\boldsymbol{\alpha}_0 + \boldsymbol{\kappa}_0^\intercal\boldsymbol{\beta}_0\\
       &+ \boldsymbol{\kappa}_2^\intercal(\boldsymbol{\alpha}_2 + \boldsymbol{\Psi}_2\boldsymbol{\alpha}_0 + \boldsymbol{\Xi}_2\boldsymbol{\alpha}_0)\\
       &+ \boldsymbol{\kappa}_2^\intercal(\boldsymbol{\beta}_2 + \boldsymbol{\Psi}_2\boldsymbol{\beta}_0 + \boldsymbol{\Xi}_2\boldsymbol{\beta}_0)]T_i\\
       &+ (\gamma_1 + \boldsymbol{\gamma}_2^\intercal\boldsymbol{\psi}_3)M_{i1}\\
       &+ (\boldsymbol{\gamma}_2^\intercal\boldsymbol{\xi}_3 + \kappa_1 + \boldsymbol{\kappa}_2^\intercal\boldsymbol{\psi}_3 + \boldsymbol{\kappa}_2^\intercal\boldsymbol{\xi}_3)M_{i1}T_i\\
       &+ (\boldsymbol{\gamma}_0^\intercal + \boldsymbol{\gamma}_2^\intercal\boldsymbol{\Psi}_2)\boldsymbol{e}_{i0} + \boldsymbol{\gamma}_2^\intercal\boldsymbol{e}_{i2}\\
       &+ (\boldsymbol{\gamma}_2^\intercal\boldsymbol{\Xi}_2 + \boldsymbol{\kappa}_0^\intercal + \boldsymbol{\kappa}_2^\intercal(\boldsymbol{\Psi}_2 + \boldsymbol{\Xi}_2))T_i\boldsymbol{e}_{i0}\\
       &+ \boldsymbol{\kappa}_2^\intercal T_i\boldsymbol{e}_{i2} + e_{i3}. \label{eq: Y}
\end{split}
\end{align}
So, we have
\begin{align*}
\theta_{M_10} ={}& \alpha_1 + \boldsymbol{\psi}_1^\intercal\boldsymbol{\alpha}_0\\
\theta_{M_11} ={}& \beta_1 + \boldsymbol{\psi}_1^\intercal\boldsymbol{\beta}_0 + \boldsymbol{\xi}_1^\intercal\boldsymbol{\alpha}_0 + \boldsymbol{\xi}_1^\intercal\boldsymbol{\beta}_0\\
\theta_{Y0} ={}& \alpha_3 + \boldsymbol{\gamma}_0^\intercal\boldsymbol{\alpha}_0 + \boldsymbol{\gamma}_2^\intercal(\boldsymbol{\alpha}_2 + \boldsymbol{\Psi}_2\boldsymbol{\alpha}_0 + \boldsymbol{\Xi}_2\boldsymbol{\alpha}_0)\\
\theta_{Y1} ={}& \beta_3 + \boldsymbol{\gamma}_0^\intercal\boldsymbol{\beta}_0 + \boldsymbol{\gamma}_2^\intercal(\boldsymbol{\beta}_2 + \boldsymbol{\Psi}_2\boldsymbol{\beta}_0 + \boldsymbol{\Xi}_2\boldsymbol{\beta}_0)\\
       &+ \boldsymbol{\kappa}_0^\intercal(\boldsymbol{\alpha}_0 + \boldsymbol{\beta}_0)\\
       &+ \boldsymbol{\kappa}_2^\intercal(\boldsymbol{\alpha}_2 + \boldsymbol{\beta}_2 + (\boldsymbol{\Psi}_2+\boldsymbol{\Xi}_2)(\boldsymbol{\alpha}_0 + \boldsymbol{\beta}_0)) \\
\theta_{Y2} ={}& \gamma_1 + \boldsymbol{\gamma}_2^\intercal\boldsymbol{\psi}_3\\
\theta_{Y3} ={}& \boldsymbol{\gamma}_2^\intercal\boldsymbol{\xi}_3 + \kappa_1 + \boldsymbol{\kappa}_2^\intercal\boldsymbol{\psi}_3 + \boldsymbol{\kappa}_2^\intercal\boldsymbol{\xi}_3\\
\mu_{M_1} ={}& (\boldsymbol{\psi}_1^\intercal + \boldsymbol{\xi}_1^\intercal  T_i)\boldsymbol{e}_{i0} + e_{i1}\\
\mu_Y ={}& (\boldsymbol{\gamma}_0^\intercal + \boldsymbol{\gamma}_2^\intercal\boldsymbol{\Psi}_2)\boldsymbol{e}_{i0} + \boldsymbol{\gamma}_2^\intercal\boldsymbol{e}_{i2}\\
       &+ (\boldsymbol{\gamma}_2^\intercal\boldsymbol{\Xi}_2 + \boldsymbol{\kappa}_0^\intercal + \boldsymbol{\kappa}_2^\intercal(\boldsymbol{\Psi}_2 + \boldsymbol{\Xi}_2))T_i\boldsymbol{e}_{i0}\\
       &+ \boldsymbol{\kappa}_2^\intercal T_i\boldsymbol{e}_{i2} + e_{i3}.
\end{align*}
Because
\begin{align*}
\mathbb{E}(\boldsymbol{e}_{i0}\vert T_i)&=\boldsymbol{0}
\\
\mathbb{E}(e_{i1}\vert T_i, \boldsymbol{M}_{i0})&=0
\\
\mathbb{E}(\boldsymbol{e}_{i2}\vert T_i, \boldsymbol{M}_{i0}, M_{i1})&=\boldsymbol{0}
\\
\mathbb{E}(e_{i3}\vert T_i, \boldsymbol{M}_{i0}, M_{i1}, \boldsymbol{M}_{i2})&=0,
\end{align*}
we have
\begin{align*}
\mathbb{E}(\mu_{M_1}\vert T_i) &= 0\\
\mathbb{E}(\mu_{Y}\vert T_i, M_{i1}) &= 0
\end{align*}
That is, \textit{strict exogeneity} conditions for Equations~\eqref{equation: lrs1} and~\eqref{equation: lrs2} are satisfied.  So their regression coefficients can be identified and consistently estimated through \textit{ordinary least squares}~\citep{Wooldridge2010EconometricData}.
\end{proof}

\section{Proof of Theorem~\ref{theorem: identification}}\label{appendix: theorem_1}
\begin{proof}
Here we show that, through substituting Equations~\eqref{eq: M1} and~\eqref{eq: Y} into definitions of {\tt  GADE} and {\tt  GACME}, {\tt  GADE} and {\tt  GACME} can be pinned down by the coefficients of the linear regression system.  Because the coefficients of the linear regression system are identified according to Lemma~\ref{lemma:lrs}, {\tt  GADE} and {\tt  GACME} are identified.
\begin{align*}
&\text{{\tt  GADE}}(t)\\
={}& \mathbb{E}[Y_i(1, \boldsymbol{M}_{i0}(1), M_{i1}(t, \boldsymbol{M}_{i0}(t)),\\
&\qquad \boldsymbol{M}_{i2}(1, \boldsymbol{M}_{i0}(1), M_{i1}(t, \boldsymbol{M}_{i0}(t))))]\\
&-\mathbb{E}[Y_i(0, \boldsymbol{M}_{i0}(0), M_{i1}(t, \boldsymbol{M}_{i0}(t)),\\
&\qquad\quad \boldsymbol{M}_{i2}(0, \boldsymbol{M}_{i0}(0), M_{i1}(t, \boldsymbol{M}_{i0}(t))))]\\
={}& \beta_3 + \boldsymbol{\gamma}_0^\intercal \mathbb{E}(\Delta\boldsymbol{M}_{i0}) + \boldsymbol{\gamma}_2^\intercal \mathbb{E}(\Delta\boldsymbol{M}_{i2})\\
   &+ \boldsymbol{\kappa}_0^\intercal \mathbb{E}[\boldsymbol{M}_{i0}(1)] + \kappa_1 \mathbb{E}[M_{i1}(t, \boldsymbol{M}_{i0}(t))]\\
   &+ \boldsymbol{\kappa}_2^\intercal \mathbb{E}[\boldsymbol{M}_{i2}(1, \boldsymbol{M}_{i0}(1), M_{i1}(t, \boldsymbol{M}_{i0}(t)))]
   \\
={}& \beta_3 + \boldsymbol{\gamma}_0^\intercal \boldsymbol{\beta}_0 + \boldsymbol{\gamma}_2^\intercal (\boldsymbol{\beta}_2 + \boldsymbol{\Psi}_2\boldsymbol{\beta}_0 + \boldsymbol{\Xi}_2\boldsymbol{\beta}_0)\\
   &+ \boldsymbol{\gamma}_2^\intercal\boldsymbol{\xi}_3(\alpha_1 + \boldsymbol{\psi}_1^\intercal\boldsymbol{\alpha}_0 + (\beta_1 + \boldsymbol{\psi}_1^\intercal\boldsymbol{\beta}_0 + \boldsymbol{\xi}_1^\intercal(\boldsymbol{\alpha}_0+\boldsymbol{\beta}_0))t)\\
   &+ \boldsymbol{\kappa}_0^\intercal (\boldsymbol{\alpha}_0 + \boldsymbol{\beta}_0)\\
   &+ \kappa_1 (\alpha_1 + \boldsymbol{\psi}_1^\intercal\boldsymbol{\alpha}_0 + (\beta_1 + \boldsymbol{\psi}_1^\intercal\boldsymbol{\beta}_0 + \boldsymbol{\xi}_1^\intercal(\boldsymbol{\alpha}_0+\boldsymbol{\beta}_0))t)\\
   &+ \boldsymbol{\kappa}_2^\intercal (\boldsymbol{\alpha}_2 + \boldsymbol{\beta}_2 + (\boldsymbol{\Psi}_2 + \boldsymbol{\Xi}_2)(\boldsymbol{\alpha}_0 + \boldsymbol{\beta}_0))\\
   &+ \boldsymbol{\kappa}_2^\intercal (\boldsymbol{\psi}_3 + \boldsymbol{\xi}_3)(\alpha_1 + \boldsymbol{\psi}_1^\intercal\boldsymbol{\alpha}_0) \\
   &+ \boldsymbol{\kappa}_2^\intercal (\boldsymbol{\psi}_3 + \boldsymbol{\xi}_3)(\beta_1 + \boldsymbol{\psi}_1^\intercal\boldsymbol{\beta}_0 + \boldsymbol{\xi}_1^\intercal(\boldsymbol{\alpha}_0+\boldsymbol{\beta}_0))t\\
={}& \theta_{Y1} + \theta_{Y3}(\theta_{M_10} + \theta_{M_11}t)
\end{align*}
\begin{align*}
&\text{{\tt  GACME}}(t)\\
={}& \mathbb{E}[Y_i(t, \boldsymbol{M}_{i0}(t), M_{i1}(1, \boldsymbol{M}_{i0}(1)),\\
&\qquad \boldsymbol{M}_{i2}(t, \boldsymbol{M}_{i0}(t), M_{i1}(1, \boldsymbol{M}_{i0}(1))))]\\
&-\mathbb{E}[Y_i(t, \boldsymbol{M}_{i0}(t), M_{i1}(0, \boldsymbol{M}_{i0}(0)),\\
&\qquad\quad \boldsymbol{M}_{i2}(t, \boldsymbol{M}_{i0}(t), M_{i1}(0, \boldsymbol{M}_{i0}(0))))]\\
={}& (\boldsymbol{\gamma}_1 + \kappa_1 t)\mathbb{E}(\Delta M_{i1}) + (\boldsymbol{\gamma}_2^\intercal +\boldsymbol{\kappa}_2^\intercal t) \mathbb{E}(\Delta\boldsymbol{M}_{i2})
   \\
={}& (\boldsymbol{\gamma}_2^\intercal +\boldsymbol{\kappa}_2^\intercal t)(\boldsymbol{\psi}_3 + \boldsymbol{\xi}_3 t)(\beta_1 + \boldsymbol{\psi}_1^\intercal\boldsymbol{\beta}_0 + \boldsymbol{\xi}_1^\intercal\boldsymbol{\alpha}_0 + \boldsymbol{\xi}_1^\intercal\boldsymbol{\beta}_0)\\
  &+ (\boldsymbol{\gamma}_1 + \kappa_1 t)(\beta_1 + \boldsymbol{\psi}_1^\intercal\boldsymbol{\beta}_0 + \boldsymbol{\xi}_1^\intercal\boldsymbol{\alpha}_0 + \boldsymbol{\xi}_1^\intercal\boldsymbol{\beta}_0)
\\
={}& \theta_{M_11}(\theta_{Y2} + \theta_{Y3}t)
\end{align*}
\end{proof}

\end{document}